\documentclass[%
aip,
amsmath,amssymb,
reprint,%
]{revtex4-1}

\usepackage{graphicx}
\usepackage{dcolumn}
\usepackage{bm}
\usepackage{amsmath}
\usepackage[utf8]{inputenc}
\usepackage[T1]{fontenc}
\usepackage{mathptmx}
\usepackage{adjustbox}
\usepackage{etoolbox}
\usepackage{algorithm}
\usepackage{xcolor}
\usepackage{enumitem}
\usepackage{float}
\usepackage{algpseudocode}    
\usepackage{capt-of}
\usepackage[hidelinks]{hyperref}

\makeatletter
\def\@email#1#2{%
	\endgroup
	\patchcmd{\titleblock@produce}
	{\frontmatter@RRAPformat}
	{\frontmatter@RRAPformat{\produce@RRAP{*#1\href{mailto:#2}{#2}}}\frontmatter@RRAPformat}
	{}{}
}%
\makeatother
\begin{document}
	
	\preprint{AIP/123-QED}
	
	\title[Punishment in bipartite societies]{Punishment in bipartite societies}
	\author{Sinan Feng}
	\affiliation{School of Mathematics and Statistics, Northwestern Polytechnical University, Xi’an, 710072, Shaanxi, China}
	\affiliation{%
		Shaanxi Provincial Key Laboratory of Intelligent Game Theory and Information Processing at Higher Education Institutions, Xi'an, 710072, China}%
	
	\author{Genjiu Xu}%
	\email{xugenjiu@nwpu.edu.cn.}
	\homepage{xugenjiu@nwpu.edu.cn.}
	\affiliation{School of Mathematics and Statistics, Northwestern Polytechnical University, Xi’an, 710072, Shaanxi, China}%
	\affiliation{Shenzhen Research Institute, Northwestern Polytechnical University, Shenzhen, 518057, Guangdong, China}%
	
	\author{Yu Chen}
	\affiliation{School of Mathematics and Statistics, Northwestern Polytechnical University, Xi’an, 710072, Shaanxi, China}
	\affiliation{%
		Shaanxi Provincial Key Laboratory of Intelligent Game Theory and Information Processing at Higher Education Institutions, Xi'an, 710072, China}%
	
	\author{Chaoqian Wang}
	\affiliation{%
		School of Mathematics and Statistics, Nanjing University of Science and Technology, Nanjing 210094, China}%
	
	\author{Attila Szolnoki}
	\affiliation{%
		Institute of Technical Physics and Materials Science, Centre for Energy Research, PO Box 49, Budapest H-1525, Hungary}
	\date{\today}
	
	\begin{abstract}
		From ant–acacia mutualism to performative conflict resolution among Inuit, dedicated punishments between distinct subsets of a population are widespread and can reshape the evolutionary trajectory of cooperation. Existing studies have focused on punishments within a homogeneous population, paying little attention to cooperative dynamics in a situation where belonging to a subset is equally important to the actual strategy represented by an actor. To fill this gap, we here study a bipartite population where cooperator agents in a public goods game penalize exclusively those defectors who belong to the alternative subset. We find that cooperation can emerge and remain stable under symmetric intergroup punishment. In particular, at low punishment intensity and at a small value of the enhancement factor of the dilemma game, intergroup punishment promotes cooperation more effectively than a uniformly applied punishment. Moreover, intergroup punishment in bipartite populations tends to be more favorable for overall social welfare. When this incentive is balanced, cooperators can collectively restrain defectors of the alternative set via aggregate interactions in a randomly formed working group, offering a more effective incentive. Conversely, breaking the symmetry of intergroup punishment inhibits cooperation, as the imbalance creates an Achilles' heel in the enforcement structure. Our work, thus, reveals symmetry in intergroup punishment as a unifying principle behind cooperation across human and biological systems.
	\end{abstract}
	
	\maketitle
	
	\begin{quotation}
		Doing bad thing is less tolerable from an external partner--this consideration is in the heart of the so-called intergroup punishment when cooperator players punish those defectors exclusively who are from an alternative subset of the population. Interestingly, the vast majority of the  punishment literature ignored this simple observation and focused on situations where all defectors are equally penalized. To reveal the potential consequence of dedicated punishment we introduce a bipartite population where players of two subsets consider each other differently. In particular, when they form a working group of public goods game, cooperators punish only those defectors who are from the alternative subset.   
		It is found that a symmetric intergroup punishment effectively facilitates the emergence 
		of cooperation. 
		In contrast, breaking the symmetry of intergroup punishment 
		undermines general cooperation by allowing defectors to exploit the less protected subset, which eventually leads to an undesired destination of the whole population.
	\end{quotation}
	
	\section{\label{sec:level1}INTRODUCTION}
	Cooperation is the basic element to maintain the operation of social, natural and economic systems~\cite{fehr2003nature,axelrod81,boyd09,perc_pr17}. Although the importance of cooperation is widely recognized, it is inherently fragile, due to the fact that participants are tempted to be self-interested, which eventually jeopardizes collective interests~\cite{hauser19}. To address this challenge, elucidating the mechanisms that generate, sustain, and enhance cooperative behavior constitutes the central pursuit of evolutionary game theory. This framework provides a foundational basis for understanding the evolution of individual strategies and the emergence of collective behavioral patterns~\cite{nowak04,nowak92,perc_jrsi13}. 
	Despite the intensive research efforts of the last decades, it has been remained essential to explain how cooperation persists in competitive and conflict environments~\cite{wang2024competition,zhu2025evolution}. 
	
	Although a large number of models assume that populations are homogeneous, real social and biological systems have significant population heterogeneity~\cite{vertebrates02,perc08,wang2022involution,wang2022reversed}. Previous studies have revealed the role of heterogeneity from different angles, indicating that population heterogeneity is complex and ubiquitous in real systems~\cite{bi23,bielawski25,li23} 
	that cannot be ignored in understanding the evolution of cooperation. Researchers have recognized that individual differences in status, abilities, and environmental conditions are key to understanding group dynamics~\cite{burlando05,corazzini24,bmc19}. Moreover, studies have shown that heterogeneity in individual attributes~\cite{hauert2025phase,wang2025inter}, network structures~\cite{huang25,wu15}, and behaviors~\cite{tsvetkova21,wei25} can substantially shape the evolution of cooperation.
	
	As a principal incentive, punishment mechanisms can further regulate cooperation levels and are, therefore, analyzed as a distinct factor in recent studies~\cite{wen25,chen14,szolnoki11,fehr02,boyd03,sigmund10,szolnoki11b,wang2024evolutionary}. Evidently, potential individual differences may have decisive role in 
	punishment mechanisms, as it was illustrated
	with attribute differences~\cite{szolnoki17,zhang22,liu19}. Battu~\cite{battu21} constructed an agent model in which ``conditional cooperation'' has different tendencies to cooperation and punishment. Xie {et al.}~\cite{xie23} investigated cooperation in public goods games under multiple heterogeneities, including contributions, punishment severity \& cost, and aspiration payoffs for strategy updating. These studies show that the effect of punishment in heterogeneous populations is much more complex than that in traditional symmetric models, and many classical conclusions do not hold in real heterogeneous environments.
	
	While the impact of punishment in heterogeneous populations has been discussed, the concerns focused on punishment within a single well-mixed population, that is, sanctions only occur among individuals with similar characteristics. In such models, the punishment relationship is limited to internal interactions and excludes more complex social structures~\cite{lu24,sigmund07}. In sharp contrast, punishment between distinct population subgroups is prevalent in real society: groups with different identities, classes, institutional positions or functional roles can sanction each other.
	
	In 1954, Hoebel~\cite{hoebel54} described that the conflict between the Inuit could be resolved by singing duels, with accusers and accusers taking turns to mock each other with poetry. Ant–acacia mutualisms illustrate a case of bidirectional punishment between distinct subgroups: plants reduce rewards to poorly defending ants, while ants retaliate by pruning low-reward plant tissues. This reciprocal enforcement prevents cheating and stabilizes cooperation~\cite{heil03}. In enterprises or cooperatives, senior managers may punish the uncooperative behavior of grassroots employees, and grassroots members may also counter management decisions through collective action~\cite{lian14}.
	
	Motivated by these real-life observations, we propose a model in which two subsets form a bipartite population. Staying at the most celebrated metaphor of conflicting self- and collective interests, we consider a public goods game where randomly selected actors from a working group for a joint venture. The basic question is whether to contribute to a common pool (cooperate) or just enjoy its fruit (defect). To avoid the undesired outcome, cooperators can penalize defector group members. In our framework, punishment occurs exclusively across subsets: cooperators penalize defectors in the other subset, and no within-group punishment is assumed. This idealized form of perfect external enforcement reflects situations commonly found in hierarchical or role-differentiated organizations, where individuals lack the capacity or legitimacy to discipline peers within their own group, while sanctioning authority operates primarily across groups.
	
	We find that symmetric punishment between distinct subsets enables cooperation to emerge and stabilize. Delicately balanced enforcement across subsets prevents any subset from becoming vulnerable to defection, thus providing a high level of collective cooperation. At relatively low punishment intensity and enhancement factor, punishment between distinct subsets promotes cooperation more effectively than a uniformly applied incentive in a homogeneous population. Cooperators of different subsets can synergistically support their independent efforts across subset divisions to collectively restrain defectors. Importantly, such cross-subset punishment in bipartite populations is also more favorable for overall social welfare. By contrast, breaking the symmetry of intergroup punishment gradually undermines cooperative behavior; hence, asymmetry creates an Achilles' heel of this incentive, by allowing defectors to exploit the less-protected subset.
	
	\section{Model}
	We consider an infinite and well-mixed bipartite population, where individuals are divided into two subsets $A$ and $B$, respectively forming $\alpha$ and ($1-\alpha$) portion of the whole population. Each individual can choose one of the basic strategies of a social dilemma: cooperation ($C$) or defection ($D$). When players are forming an $N$-member working group, cooperators incur a contribution cost of $c=1$ to promote the public good. Defectors, in contrast, do not contribute to the common pool, but only enjoy its fruit.
	
	To avoid the undesired tragedy of the common state, cooperators may impose peer punishment~\cite{fehr02} on defectors at their own expense. There is, however, a significant difference from the intensively studied scenario in which {\it all} defectors are punished by cooperator players. Instead, we assume that a cooperator player considers not only the actual strategy of a partner but also the subset from which a defector is from. More precisely, a cooperator punishes only those defectors who are members of the opposite subset.  
	
	This perfect external monitoring captures key dynamics of cross-group punishment without aiming to describe all real-world sanctioning mechanisms. In many institutional or organizational settings characterized by role differentiation or power asymmetries, individuals lack the capacity or legitimacy to sanction peers within their own subset, while sanctioning authority operates primarily across groups. By focusing on this limiting case, we abstract from within group enforcement and emphasize the role of cross subset punishment in shaping cooperative dynamics in a transparent and analytically tractable manner.
	
	Before discussing the potential consequences of this dedicated, or intergroup punishment, we first briefly summarize the driving rule of the basic model where players are uniform and belong to the same set. Accordingly, cooperators penalize all defectors in the working group on their own expenses.
	
	\subsection{Uniform population}
	
	In a uniform population, the only tag of a player is its strategy. Accordingly, the
	payoff functions for cooperators ($\pi_C$) and defectors ($\pi_D$) 
	are formulated as follows:
	
	\begin{equation}
		\begin{cases}
			\pi_C = \displaystyle{\frac{r\,(n_C+1)}{N} - 1 - n_D\lambda}\\[8pt]
			\pi_D = \displaystyle{\frac{r\,n_C}{N} - n_C\gamma},
		\end{cases}
	\end{equation}
	where $ N $ denotes the group size in a single game and $r$ is the enhancement factor amplifying cooperative contributions. $n_C$ and $n_D$ represent the number of cooperators and defectors in the single game. Furthermore, $\lambda$ and $\gamma$ denote the punishment cost and fine imposed on defectors in the single game.
	
	Accordingly, the expected payoffs of cooperators ($f_C$) and defectors ($f_D$) within the group are given by:
	\begin{equation}
		\begin{cases}
			f_C 
			&=\displaystyle{ \sum_{n_C + n_D = N - 1} 
				\frac{(N-1)!}{n_C! \, n_D!} 
				\left(x_C\right)^{n_C} 
				\left(x_D\right)^{n_D} \pi_C} \\[0.3em]
			&= \displaystyle{\frac{r}{N} (N-1)x_C + \frac{r}{N} - 1 - (N-1) \lambda (1-x_C)}\\[0.8em]
			f_D
			&= \displaystyle{\sum_{n_C + n_D = N - 1} 
				\frac{(N-1)!}{n_C! \, n_D!} 
				\left(x_C\right)^{n_C} 
				\left(x_D\right)^{n_D} \pi_D} \\[0.3em]
			&=
			\displaystyle{\frac{r}{N} (N-1)x_C - \gamma (N-1)x_C},
		\end{cases}
	\end{equation}
	where $x_C$ and $x_D$ denote the proportions of cooperators and defectors, respectively. Based on the replicator dynamics theory, the evolutionary equation of the cooperation strategy is expressed as follows:
	\begin{equation}
		\begin{aligned}
			\dot{x}_C 
			&= x_C \bigl(1 - x_C\bigr)\bigl(f_C - f_D\bigr) \\
			&=\displaystyle{ x_C \bigl(1 - x_C\bigr)
				\left[\frac{r}{N} - 1 - \lambda (N - 1)x_D 
				+ \gamma (N - 1)x_C \right]}.
		\end{aligned}
	\end{equation}
	
	\subsection{Bipartite population}
	
\begin{figure}[htbp]
		\hspace*{-0.2cm}
		
		\adjustbox{rotate=90, max width=0.3\textheight, center}{\includegraphics{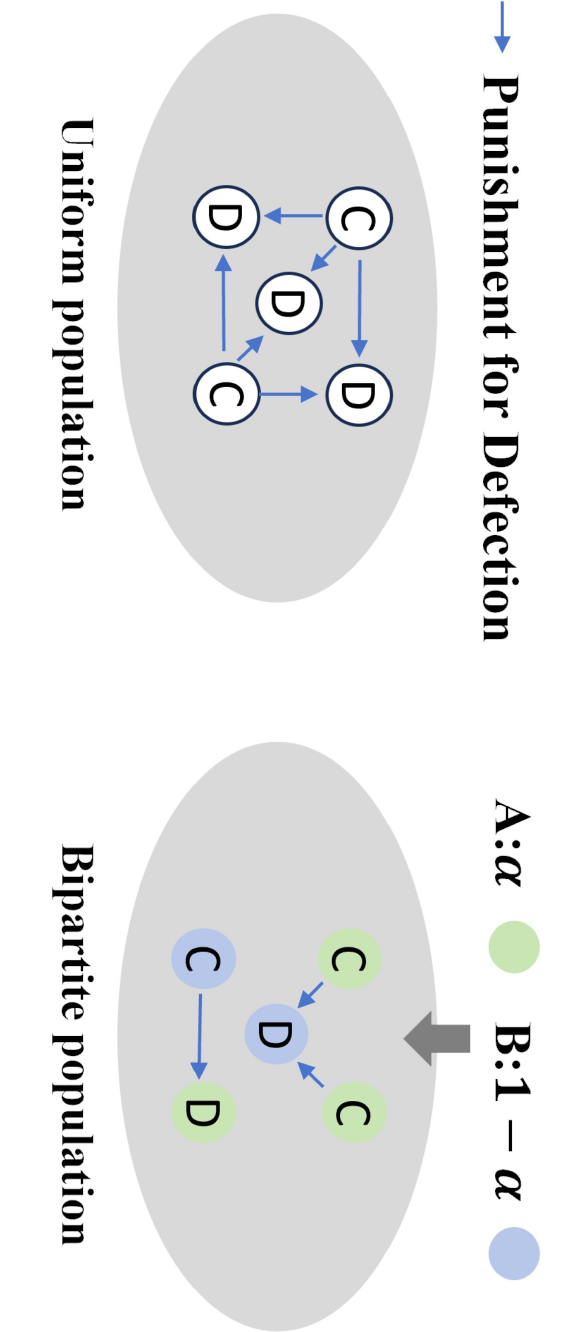}}
		\caption{\label{fig:1} Illustration of punishment mechanisms within a uniform population and a bipartite population in the framework of public goods game. (a) In a uniform population, cooperators punish all defectors within their working group. (b) In a bipartite population, cooperators penalize only those defectors who are from the alternative subsets. Green and blue colors mark the different subsets and arrows denote the actual punishment act between group members.}
	\end{figure}

	In the alternative model, which is the key part of our present work, we assume a bipartite population where players belong to two subsets and cooperators punish wrongdoers differently. Importantly, this bipartition is purely a labeling of individuals; group formation for the public goods game remains well-mixed, so that any individual can interact with any other within the working group. In the simplified version, a cooperator punishes only those defectors who are from the alternative set. The key differences between the uniform and bipartite populations are summarized in Fig.~\ref{fig:1}.
	In real-world scenarios, individuals 
	are frequently organized in different sets
	and punishment usually interacts between departments within the organization, teams within the community, or social groups within the society~\cite{tepper17}. 
	
	A common example is a working mode of interaction between leaders and followers. When leaders ignore rules or behave unfairly, followers often punish leaders by postponing tasks, reducing efforts, or reporting upward. Conversely, when followers violate expectations, leaders may discipline them or shelve opportunities. Therefore, these two groups are mutually regulated, rather than relying solely on the discipline within their respective groups. Interestingly, it is often observed that when either party's supervision of the other is roughly balanced, both groups tend to act more responsibly. This kind of mutual regulation between distinct groups is generally considered to have a stronger impact than when it is regulated only within the leader group or within the follower group, because both parties care about how the other party evaluates and responds to their behavior.
	
	Inspired by these examples, we consider the punishment model between distinct subsets, where punishment occurs across different subsets, a mechanism often neglected in previous studies. Specifically, cooperators of subset $A$ impose punishment on defectors of subset $B$, and similarly, subset-$B$ cooperators punish subset-$A$ defectors. As noted, the proportions of subsets are denoted by $\alpha$ and $1-\alpha$, respectively. The payoff functions for each strategy can be expressed as
	\begin{equation}
		\left\{
		\begin{aligned}
			\pi_{C}^{[A]} &= \frac{r\bigl(n_{C}^{[A]} + n_{C}^{[B]} + 1\bigr)}{N} - 1 - n_{D}^{[B]}\lambda_{A} \\
			\pi_{D}^{[A]} &= \frac{r\bigl(n_{C}^{[A]} + n_{C}^{[B]}\bigr)}{N} - n_{C}^{[B]}\gamma_{B} \\
			\pi_{C}^{[B]} &= \frac{r\bigl(n_{C}^{[A]} + n_{C}^{[B]} + 1\bigr)}{N} - 1 - n_{D}^{[A]}\lambda_{B} \\
			\pi_{D}^{[B]} &= \frac{r\bigl(n_{C}^{[A]} + n_{C}^{[B]}\bigr)}{N} - n_{C}^{[A]}\gamma_{A},
		\end{aligned}
		\right.
	\end{equation}
	where $N$ is the group size in a single game. $\gamma_A$ is the punishment fine imposed by subset-$A$ cooperators on subset-$B$ defectors, while $\gamma_B$ is the punishment fine imposed by subset-$B$ cooperators on subset-$A$ defectors.
	
	Therefore, the expected payoff of each combination strategy can be obtained as 
	
	\begin{widetext}
		\begin{equation}
			\left\{
			\begin{aligned}
				f_{C}^{[A]} &= 
				\sum_{\substack{n_{C}^{[A]} + n_{D}^{[A]} + n_{C}^{[B]} + n_{D}^{[B]} = N-1}}
				\frac{(N-1)!}{n_{C}^{[A]}! \, n_{D}^{[A]}! \, n_{C}^{[B]}! \, n_{D}^{[B]}!} \,
				\bigl(x_{C}^{[A]}\bigr)^{n_{C}^{[A]}}
				\bigl(\alpha - x_{C}^{[A]}\bigr)^{n_{D}^{[A]}}
				\bigl(x_{C}^{[B]}\bigr)^{n_{C}^{[B]}}
				\bigl(1-\alpha - x_{C}^{[B]}\bigr)^{n_{D}^{[B]}}
				\, \pi_{C}^{[A]} \\
				&= \frac{r}{N}(N-1)\bigl(x_{C}^{[A]} + x_{C}^{[B]}\bigr) + \frac{r}{N} - 1 - (N-1)\lambda_{A}\bigl(1-\alpha - x_{C}^{[B]}\bigr) \\[1.2em]
				f_{D}^{[A]} &= 
				\sum_{\substack{n_{C}^{[A]} + n_{D}^{[A]} + n_{C}^{[B]} + n_{D}^{[B]} = N-1}}
				\frac{(N-1)!}{n_{C}^{[A]}! \, n_{D}^{[A]}! \, n_{C}^{[B]}! \, n_{D}^{[B]}!} \,
				\bigl(x_{C}^{[A]}\bigr)^{n_{C}^{[A]}}
				\bigl(\alpha - x_{C}^{[A]}\bigr)^{n_{D}^{[A]}}
				\bigl(x_{C}^{[B]}\bigr)^{n_{C}^{[B]}}
				\bigl(1-\alpha - x_{C}^{[B]}\bigr)^{n_{D}^{[B]}}
				\, \pi_{D}^{[A]} \\
				&= \frac{r}{N}(N-1)\bigl(x_{C}^{[A]} + x_{C}^{[B]}\bigr) - (N-1)\gamma_{B}x_{C}^{[B]} \\[1.2em]
				f_{C}^{[B]} &= 
				\sum_{\substack{n_{C}^{[A]} + n_{D}^{[A]} + n_{C}^{[B]} + n_{D}^{[B]} = N-1}}
				\frac{(N-1)!}{n_{C}^{[A]}! \, n_{D}^{[A]}! \, n_{C}^{[B]}! \, n_{D}^{[B]}!} \,
				\bigl(x_{C}^{[A]}\bigr)^{n_{C}^{[A]}}
				\bigl(\alpha - x_{C}^{[A]}\bigr)^{n_{D}^{[A]}}
				\bigl(x_{C}^{[B]}\bigr)^{n_{C}^{[B]}}
				\bigl(1-\alpha - x_{C}^{[B]}\bigr)^{n_{D}^{[B]}}
				\, \pi_{C}^{[B]} \\
				&= \frac{r}{N}(N-1)\bigl(x_{C}^{[A]} + x_{C}^{[B]}\bigr) + \frac{r}{N} - 1 - (N-1)\lambda_{B}\bigl(\alpha - x_{C}^{[A]}\bigr) \\[1.2em]
				f_{D}^{[B]} &= 
				\sum_{\substack{n_{C}^{[A]} + n_{D}^{[A]} + n_{C}^{[B]} + n_{D}^{[B]} = N-1}}
				\frac{(N-1)!}{n_{C}^{[A]}! \, n_{D}^{[A]}! \, n_{C}^{[B]}! \, n_{D}^{[B]}!} \,
				\bigl(x_{C}^{[A]}\bigr)^{n_{C}^{[A]}}
				\bigl(\alpha - x_{C}^{[A]}\bigr)^{n_{D}^{[A]}}
				\bigl(x_{C}^{[B]}\bigr)^{n_{C}^{[B]}}
				\bigl(1-\alpha - x_{C}^{[B]}\bigr)^{n_{D}^{[B]}}
				\, \pi_{D}^{[B]} \\
				&= \frac{r}{N}(N-1)\bigl(x_{C}^{[A]} + x_{C}^{[B]}\bigr) - (N-1)\gamma_{A}x_{C}^{[A]}.
			\end{aligned}
			\right.
		\end{equation}
	\end{widetext}
	
	The corresponding replicator equations are
	
	\begin{widetext}
		\begin{equation}
			\left\{
			\begin{aligned}
				\dot{x}_{C}^{[A]} 
				&= x_{C}^{[A]}\bigl(\alpha - x_{C}^{[A]}\bigr)\bigl(f_{C}^{[A]} - f_{D}^{[A]}\bigr)  x_{C}^{[A]}\bigl(\alpha - x_{C}^{[A]}\bigr)
				\Bigl[
				\tfrac{r}{N} - 1 
				- (N - 1)\lambda_{A}\bigl(1 - \alpha - x_{C}^{[B]}\bigr)
				+ (N - 1)\gamma_{B}x_{C}^{[B]}
				\Bigr]\\[1em]
				\dot{x}_{C}^{[B]} 
				&= x_{C}^{[B]}\bigl(1 - \alpha - x_{C}^{[B]}\bigr)\bigl(f_{C}^{[B]} - f_{D}^{[B]}\bigr)  x_{C}^{[B]}\bigl(1 - \alpha - x_{C}^{[B]}\bigr)
				\Bigl[
				\tfrac{r}{N} - 1 
				- (N - 1)\lambda_{B}\bigl(\alpha - x_{C}^{[A]}\bigr)
				+ (N - 1)\gamma_{A}x_{C}^{[A]}
				\Bigr].
			\end{aligned}
			\right.
		\end{equation}
	\end{widetext}
	
	\section{Results}
	\subsection{Uniform population}
	We first present the basic model in which cooperators do not distinguish defectors in the working group and punish them all. Under this framework, the evolutionary dynamics are described by Eq.~(3). Solving this equation yields three equilibrium points: the full-defection state $x_C=0$, the full-cooperation state $x_C=1$, and an interior equilibrium $x_C^*=\displaystyle{\frac{(N - 1)\lambda + 1 - r/N}{(N - 1)(\gamma + \lambda)}}$.
	
	To analyze the evolutionary dynamics, we consider the replicator equation $\dot{x}_C = x_C (1 - x_C) f(x_C)$, where $f(x_C)$ represents the payoff difference between cooperators and defectors. 
	Since the derivative $f'(x_C) = \gamma (N-1) > 0$, the payoff difference increases monotonically with $x_C$, indicating that stronger punishment $\gamma$ enhances the growth rate of cooperators. In particular, larger $\gamma$ shifts the internal equilibrium $x_C^*$ to lower values, enlarging the basin of attraction of the fully cooperative state by making defection less favorable. 
	
\begin{figure*}[htbp]
		\adjustbox{rotate=90, max width=0.6\textheight, center}{
			\includegraphics[trim=1cm 0cm 1cm 0cm, clip]{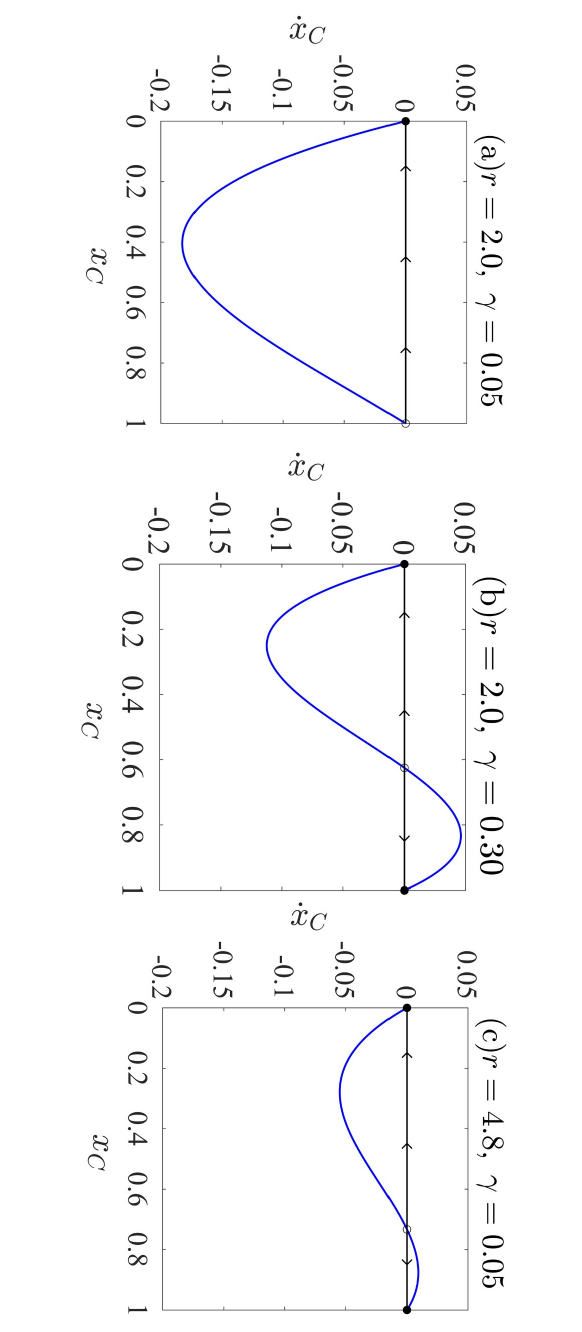}
		}
		\caption{\label{fig:2} Evolutionary dynamics in infinite mixed subgroups. Panels~(a)-(c) show ${\dot{x}_C}$ as a function of ${x_{C}}$ for different enhancement factors $r$ and punishment fine $\gamma$. Solid lines indicate the dynamics, arrows show the direction of evolution, and filled circles mark stable equilibria, while open circles mark unstable equilibria. The parameters are set as $N=5, \lambda=0.1$, (a) $r=2, \gamma=0.05$, (b) $r=2, \gamma=0.3$, (c) $r=4.8, \gamma=0.05$.}
	\end{figure*}

	When the enhancement factor satisfies $r < N [1 + \lambda (N-1)]$, the payoff difference near zero is negative, $\lim_{x_C \to 0^+} f(x_C) < 0$, leading to $\dot{x}_C < 0$ and rendering the boundary equilibrium $x_C = 0$ stable, as shown in Fig.~\ref{fig:2}(a). In this regime, the benefits of cooperation are insufficient to overcome the temptation to defect, preventing cooperators from spreading. This stability persists for all admissible parameter values if $\gamma$ is sufficiently small. 
	
	By contrast, if either the enhancement factor $r$ or the punishment intensity $\gamma$ is sufficiently large, the system exhibits bistability. In this case, $\lim_{x_C \to 1^-} f(x_C) > 0$, implying $\dot{x}_C > 0$ and making the boundary equilibrium $x_C = 1$ stable. Within this bistable regime, the system also possesses an internal unstable equilibrium at $\displaystyle{x_C^* = \frac{(N-1)\lambda + 1 - r/N}{(N-1)(\gamma + \lambda)}}$, as illustrated in Fig.~\ref{fig:2}(b) and (c). At high $r$, cooperative benefits outweigh defection, enabling cooperators to spread, while stronger punishment reinforces this effect. In general, Fig.~\ref{fig:2} shows that increasing either the punishment intensity $\gamma$ or the enhancement factor $r$ promotes cooperation.
	
	\begin{figure*}[htbp]
		\adjustbox{rotate=90, max width=0.6\textheight, center}{
			\includegraphics[trim=0cm 3cm 0cm 3cm, clip]{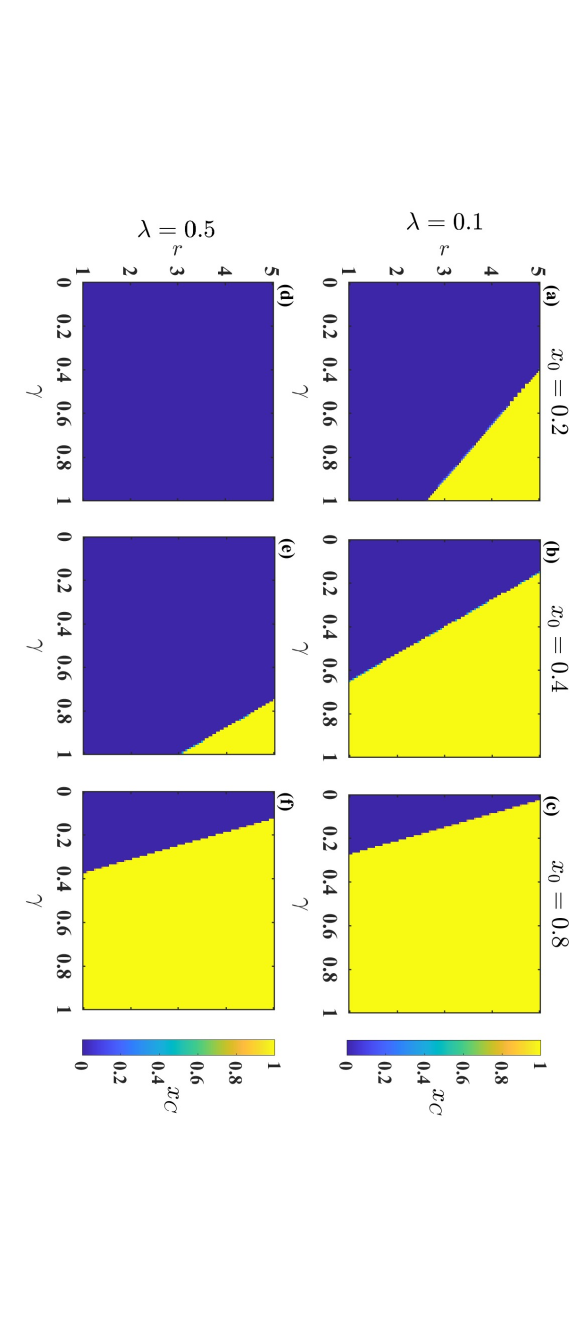}
		}
		\caption{\label{fig:3} Stable distribution of the cooperation level ${x}_C$ under different initial conditions $x_0$ and punishment fine $\lambda$. Panels~(a–c) correspond to $\lambda=0.1$, and panels~(d–f) correspond to $\lambda=0.5$, with initial cooperation levels $x_0=0.2,0.4,0.8$, respectively. The group size is $N=5$ in all panels.}
	\end{figure*}
	
	To further examine the impact of uniform punishment, Fig.~\ref{fig:3} shows the cooperator fraction $x_C$ under varying $\gamma$ and $r$ at two representative $\lambda$ values. The upper and lower rows correspond to different punishment costs $\lambda$, while the columns show varying initial cooperator fractions $x_0$, illustrating their combined influence on the evolution of cooperation.
	
	As shown in Fig.~\ref{fig:3}, with increasing $\gamma$, the fraction of cooperators ${x}_C$ generally increases, since stronger punishment effectively reduces the payoff advantage of defectors and thereby promotes the spread and stability of cooperative behavior. Conversely, when $\gamma$ is small, defectors gain higher relative payoffs, leading the population to evolve toward full defection. Similarly, a larger enhancement factor $r$ amplifies the collective benefit generated by cooperators, improving the relative payoff of cooperative groups and, thus, facilitating the maintenance of cooperation. Therefore, the transition from defection to cooperation observed in the diagram arises from the synergistic effect of increasing both the punishment fine and the enhancement factor, which together enlarge the parameter region where cooperation becomes stable.
	
	As the initial proportion of cooperators $x_0$ increases, the system tends to evolve toward full cooperation under the same punishment cost $\lambda$. In a well-mixed population, whether cooperation succeeds depends on whether $x_0$ exceeds the internal unstable equilibrium ${x_C}^*$, which determines if cooperators can invade defectors. As $r$ and $\gamma$ increase, the system enters a bistable regime. In this regime, if the initial fraction $x_0$ of cooperators exceeds the internal equilibrium point ${x_C}^*$ (Fig.~\ref{fig:3}(a) as an example, when $r=4, \gamma=0.8$, ${x_C}^*=1/6<x_0$), the system will eventually evolve toward a full cooperation state, as illustrated in Fig.~\ref{fig:3}(a-c), (e), and (f). On the contrary, a low $\lambda$ allows cooperation to arise even for moderate values of $r$ and $\gamma_A$. As $\lambda$ increases, ${x_C}^*$ rises because higher punishment costs reduce cooperators’ payoff, weakening their advantage over defectors and shrinking the cooperative region. Stable cooperation then requires higher $x_0$, stronger $\gamma$, or larger $r$. In sum, under uniform punishment protocol, cooperation emerges depending on whether the initial cooperator fraction exceeds the internal equilibrium and if the incentive can offset the punishment cost.
	
	\subsection{Bipartite population}
	We now present the results obtained in a bipartite population where punishment is restricted to partners from the alternative subset.
	In this case, the evolutionary dynamics of cooperation is described by Eq.~(6). By solving this equation, four boundary equilibrium $(0,0), (0,1-\alpha), (\alpha,0), (\alpha,1-\alpha)$ and one interior equilibrium $({x_C^{[A]}}^*,{x_C^{[B]}}^*)$ can be obtained. Among them, only $(0,0)$ remains stable under all conditions, whereas the equilibrium $(\alpha,1-\alpha)$ is stable only when $r>\max\Biggl\{N\Big[1-(N-1)\gamma_B(1-\alpha)\Big],N\Big[1-(N-1)\gamma_A\alpha\Big]\Biggr\}$. This implies that the system may evolve either into a fully defective state or exhibit bistability. Additionally, both symmetric and asymmetric cases of punishment between distinct subgroups are analyzed.
	
	\subsubsection{Symmetrical punishment}
	Under symmetric punishment between distinct subsets, the two sub-population have identical payoff structures and, thus, follow the same evolutionary dynamics. In this setting, cooperators in both sets bear the same punishment cost and impose the same punishment fine to the members of the opposite subset, ensuring complete symmetry in behavioral responses.
	
	\begin{figure*}
		\hspace*{-1cm}
		\includegraphics[width=0.8\linewidth]{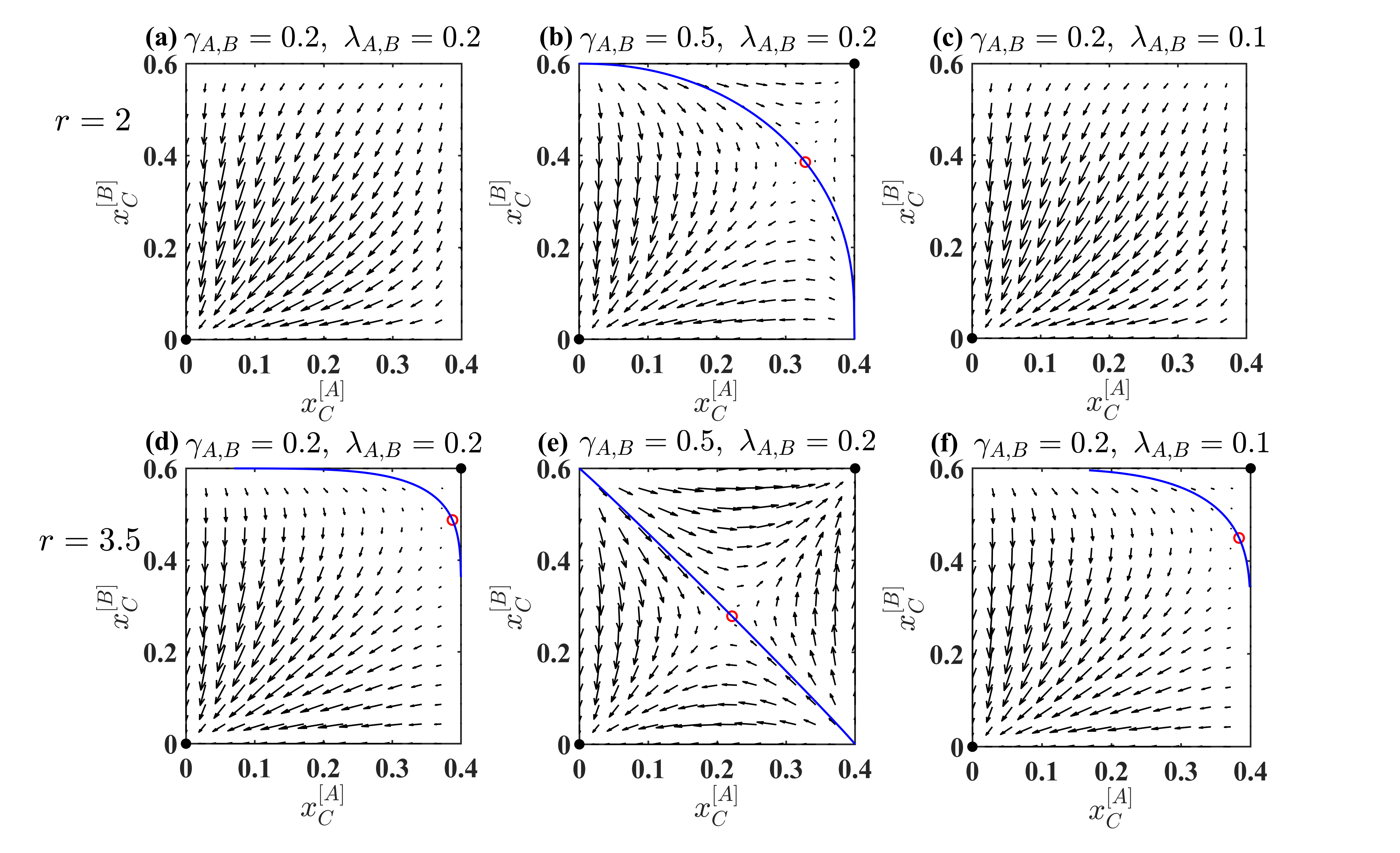}
		\caption{\label{fig:4} Phase flow of subset-$A$ and subset-$B$  cooperators under intergroup punishment. Arrows show evolutionary trajectories in the ${x_C^{[A]}},{x_C^{[B]}}$ plane. Blue curves denote stable manifolds separating two basins of attraction. Red open circles mark saddle points, and black closed circles mark stable equilibria. Panels~(a–c) correspond to $r=2$, and (d–f) to $r=3.5$. The punishment fine and cost vary as $(\gamma_{A,B}, \lambda_{A,B})=(0.2, 0.2), (0.5,0.2), (0.2,0.1)$, respectively.}
	\end{figure*}
	
	Fig.~\ref{fig:4} presents the phase portraits of the replicator dynamics under the punishment mechanism between distinct subsets for different parameter combinations. The stable manifold geometrically characterizes the boundary between different basins of attraction in the system. The stable manifold of a saddle point $x^*$ is defined as:
	\begin{equation}
		W_s(\mathbf{x}^*) = \left\{ \mathbf{x} \in \mathbb{R}^2 : \lim_{t \to \infty} \varphi(t; \mathbf{x}) = \mathbf{x}^* \right\}.
	\end{equation}
	
	In numerical implementation, the stable manifold can be efficiently approximated using the eigenvector method. After locating the saddle point and computing its Jacobian $J$, the eigenvector $v_s$ associated with the negative real eigenvalue is identified as the local stable direction. Small perturbations are then applied along $\pm v_s$, and by numerically integrating the corresponding ordinary differential equations, the one-dimensional stable manifold can be obtained. The overall computational procedure is summarized in the following pseudocode:
	
	\begin{algorithm}[H]
		\caption{Stable Manifold Construction near a Saddle Point}
		\label{alg:stable-manifold}
		\begin{algorithmic}[1]
			\Require Vector field $f(\mathbf{x}, \mu)$, saddle point $\mathbf{x}^*$,
			finite difference step $h$, time step $\Delta t$, perturbation amplitude $\Delta$
			\Ensure Stable manifold point set $M = \{\mathbf{x}(t)\}$
			
			\State Compute the Jacobian $J$ at $\mathbf{x}^*$ using central differences.
			\State Obtain the eigenpair $(\lambda_s, \mathbf{v}_s)$ of $J$ with $\mathrm{Re}(\lambda_s) < 0$ (stable direction).
			\State Initialize perturbed points: 
			$\mathbf{x}_0^{+} = \mathbf{x}^* + \Delta \mathbf{v}_s$, 
			$\mathbf{x}_0^{-} = \mathbf{x}^* - \Delta \mathbf{v}_s$.
			\For{each $\mathbf{x}_0 \in \{\mathbf{x}_0^{+}, \mathbf{x}_0^{-}\}$}
			\For{$k = 1$ to $N_{\max}$}
			\State $\mathbf{x} \gets \mathrm{RK4Step}(\mathbf{x}, \Delta t, -f)$ \Comment{Integrate the reversed system $\dot{\mathbf{x}} = -f(\mathbf{x})$}
			\If{$\mathbf{x}$ leaves domain or approaches another equilibrium}
			\State \textbf{break}
			\EndIf
			\State Record $\mathbf{x}$ into manifold set $M$
			\EndFor
			\EndFor
			\State \Return $M$
		\end{algorithmic}
	\end{algorithm}
	
	The impact of key parameters on the system’s evolutionary dynamics is clearly reflected in Fig.~\ref{fig:4}. Increasing the enhancement factor $r$ (comparing panels~(a–c) with (d–f)) raises the relative payoff of cooperation and noticeably expands the basin of attraction leading to full cooperation. Consequently, the stable manifold shifts toward the full-defection region, promoting a higher overall cooperation level. Strengthening the punishment fine $\gamma_{A,B}$ (as seen when comparing Fig.~\ref{fig:4}(a) with (b), and (d) with (e)) further enlarges the region favoring cooperation, because stronger punishment between distinct subsets increases the cost of defection and improves the relative payoff of cooperators. Likewise, a reduction in punishment cost $\lambda_{A,B}$ (comparisons between Fig.~\ref{fig:4}(b) and (c) and between (e) and (f)) reinforces cooperative stability. When punishment becomes cheaper, cooperators can afford to enforce prosocial behavior more frequently, sustaining cooperation without being drained by enforcement costs. When punishment is expensive, however, cooperators are forced to punish sparingly, eroding their strategic advantage and allowing defectors to expand. Therefore, large punishment costs weaken cooperative behavior by shrinking the cooperative region and enabling the defectors’ basin of attraction to dominate. In sum, high $r$ and $\gamma_{A,B}$, along with low $\lambda_{A,B}$, facilitate the emergence of a stable double-cooperation equilibrium.
	
\begin{figure*}[htbp]
		\adjustbox{rotate=90, max width=0.5\textheight, center}{
			\includegraphics[trim=0cm 0cm 0cm 0cm, clip]{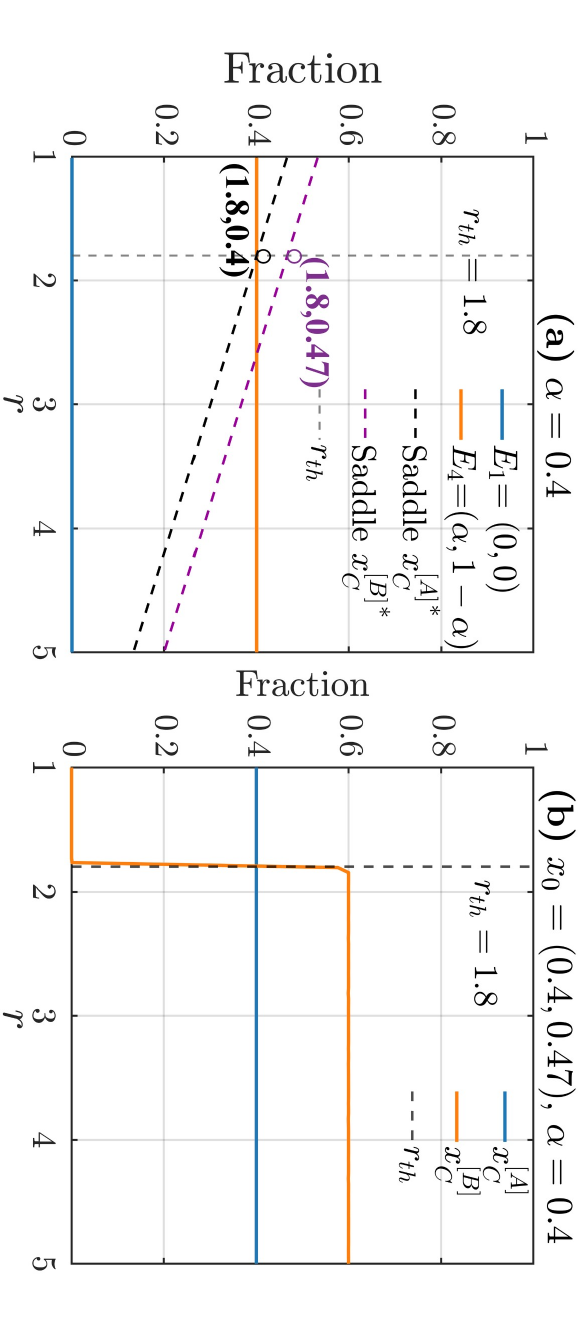}
		}
		\caption{\label{fig:5} Saddle point curves and cooperator fractions of subset-$A$ and subset-$B$ under symmetric punishment between subsets, plotted as functions of the multiplication factor $r$. Panel~(a) illustrates the saddle points $x{_C^{[A]}}^*$ and $x{_C^{[B]}}^*$ as $r$ varies. $E_1=(0,0)$ and $E_4=(\alpha,1-\alpha)$ are two stable equilibria. Panel~(b) shows the fractions of cooperators $x_C^{[A]}$ and $x_C^{[B]}$. The vertical dashed line marks the threshold $r_{th}$, above which a stable internal equilibrium emerges. Parameters: (b) $x_0=(0.4,0.47)$, $\alpha=0.4$, $\lambda_{A,B}=0.2$, $\gamma_{A,B}=0.4$ and $N=5$.}
	\end{figure*}

	We systematically explore how the interplay of critical parameters under the symmetric punishment mechanism between distinct subsets shapes the equilibrium landscape and facilitates the establishment of stable cooperation. Fig.~\ref{fig:5} analyzes the effect of varying $r$ on the evolution of cooperative behavior. When $r>r_{th}(r_{th}=\max\{N[1-(N-1)\gamma_B(1-\alpha)],\,N[1-(N-1)\gamma_A\alpha]\})$, the internal equilibrium becomes stable, indicating that the system can theoretically reach a fully cooperative stable state. However, the coexistence of bistability means that total defection may also occur. In the bistable regime, which state the system ultimately reaches is determined by the saddle point at $r_{th}$ and relative to the initial cooperator proportion. 
	
	Panel~(a) of Fig.~\ref{fig:5} shows the evolution of the saddle points $x{_C^{[A]}}^*$ and $x{_C^{[B]}}^*$ as $r$ increases, in comparison to the two stable equilibria. Since the increase of $r$ will increase the payoffs of cooperators, the saddle point curves of the two subsets for cooperators decrease monotonously with the increase of $r$, indicating that the increase of $r$ is beneficial to the system to enter the cooperative state, as shown in Fig.~\ref{fig:5}(a). This pattern reflects a key mechanism of symmetric punishment between distinct subsets: both subsets impose and receive punishment at equivalent strength and cost, so an increase in $r$ amplifies cooperative gains uniformly in both subsets, allowing neither side to become a liability nor a burden. Essentially, symmetry prevents vulnerable points from forming, allowing cooperative reinforcement to diffuse smoothly across the bipartite system.
	
	The saddle point $(0.4,0.47)$ at the threshold $r_{th}=1.8$ is taken as the initial fraction of cooperators for studying the evolutionary dynamics under varying $r$ as shown in Fig.~\ref{fig:5}(b). Since the initial value $x_0=0.4$ of subgroup-$A$ cooperators equals their stable state value, they remain stable from the beginning. This reflects the mechanism that when cooperators start at or above their internal equilibrium, they are self-sustaining in a well-mixed population. In contrast, subset-$B$ cooperators reach stability only when $r>r_{th}$. Collectively, these results indicate that increasing $r$ not only promotes the stability of internal equilibria but also expands the basin of attraction for cooperative states.
	
	\subsubsection{Asymmetric punishment}
	The following analysis considers an asymmetric punishment mechanism between distinct subsets. Specifically, cooperators in the subset-A incur a relatively lower cost $\lambda_A$ when punishing defectors in the subset-B, whereas cooperators in the subset-B bear a higher cost $\lambda_B$ ($\lambda_A<\lambda_B$) when punishing defectors of subset-A. Meanwhile, the punishment intensity exerted by subset-A cooperators on subset-B defectors $\gamma_A$ is stronger than that imposed by subset-B cooperators on subset-A defectors $\gamma_B$ ($\gamma_A>\gamma_B$).
	
	\begin{figure*}[htbp]
		\hspace*{-0.4cm}
		\adjustbox{rotate=90, max width=0.5\textheight, center}{
			\includegraphics[trim=0cm 6cm 0cm 5cm, clip]{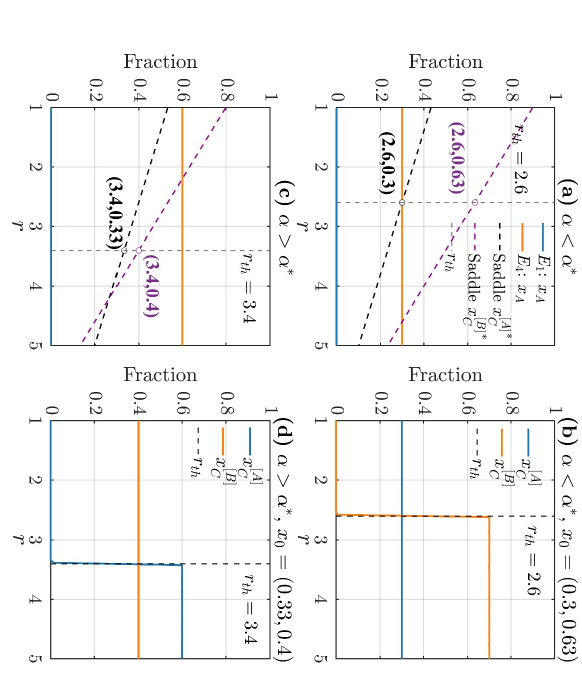}
		}
		\caption{\label{fig:6} Saddle point curves and cooperator fractions of subset-$A$ and subset-$B$ under asymmetric punishment between distinct subsets, plotted as functions of the multiplication factor $r$. The top and bottom rows correspond to $\alpha=0.3$ and $\alpha=0.6$, respectively. Panels~(a) and (c) illustrate the saddle points $x{_C^{[A]}}^*$ and $x{_C^{[B]}}^*$ as $r$ varies. $E_1=(0,0)$ and $E_4=(\alpha,1-\alpha)$ are two stable equilibria. Panels~(b) and (d) show the fractions of cooperators $x_C^{[A]}$ and $x_C^{[B]}$. The vertical dashed line marks the threshold $r_{th}$ where a stable internal equilibrium emerges. Parameters are (b) $x_0=(0.3,0.63)$ and (d) $x_0=(0.33,0.4)$. Other parameters are $\lambda_A=0.1$, $\lambda_B=0.2$, $\gamma_A=0.4$, $\gamma_B=0.2$ and $N=5$.}
	\end{figure*}
	
	Under asymmetric punishment between distinct subsets, the stability of the internal equilibrium depends on the value of $\alpha$. When $\alpha < \alpha^* = \gamma_B / (\gamma_A + \gamma_B)$, the equilibrium is stable only if $r > N\big[1 - (N-1)\gamma_A \alpha \big]$. However, when $\alpha >\alpha^*$, stability is achieved only if $r > N\big[1 - (N-1)\gamma_B (1-\alpha) \big]$. This threshold division reflects the shifting balance of enforcement power: at low $\alpha$, dominant subset-A cooperators primarily control the promotion of cooperation, whereas at high $\alpha$, the influence of subset-B cooperators becomes increasingly important. Fig.~\ref{fig:6} illustrates how cooperation evolves with the multiplication factor $r$ in these two cases.
	
	As shown in Fig.~\ref{fig:6}, increasing $r$ promotes the evolution of cooperation within the system. When $\alpha$ is small ($\alpha = 0.3$), the threshold for achieving full cooperation is primarily determined by the punishment intensity $\gamma_A$ exerted by the dominant subset-A individuals. Therefore, as shown in Fig.~\ref{fig:6}(a) and Fig.~\ref{fig:6}(b), even relatively small values of $r$ are sufficient to stabilize full cooperation. This occurs because subset-$A$ individuals, being dominant at low $\alpha$, can effectively project cooperative influence across subgroups, suppressing defection efficiently and reducing the collective incentive needed to reach full cooperation. In contrast, subset-$B$ cooperators, initially constrained by higher punishment costs and weaker enforcement power, require higher $r$ to attain the same cooperative stability. Under this mechanism, the contribution of subset-$A$ individuals to promoting cooperation consistently exceeds that of subset-$B$ individuals.
	
	\begin{figure*}[htbp]
		\hspace*{-0.2cm}
		\adjustbox{rotate=90, max width=0.6\textheight, center}{
			\includegraphics[trim=1cm 0cm 1cm 0cm, clip]{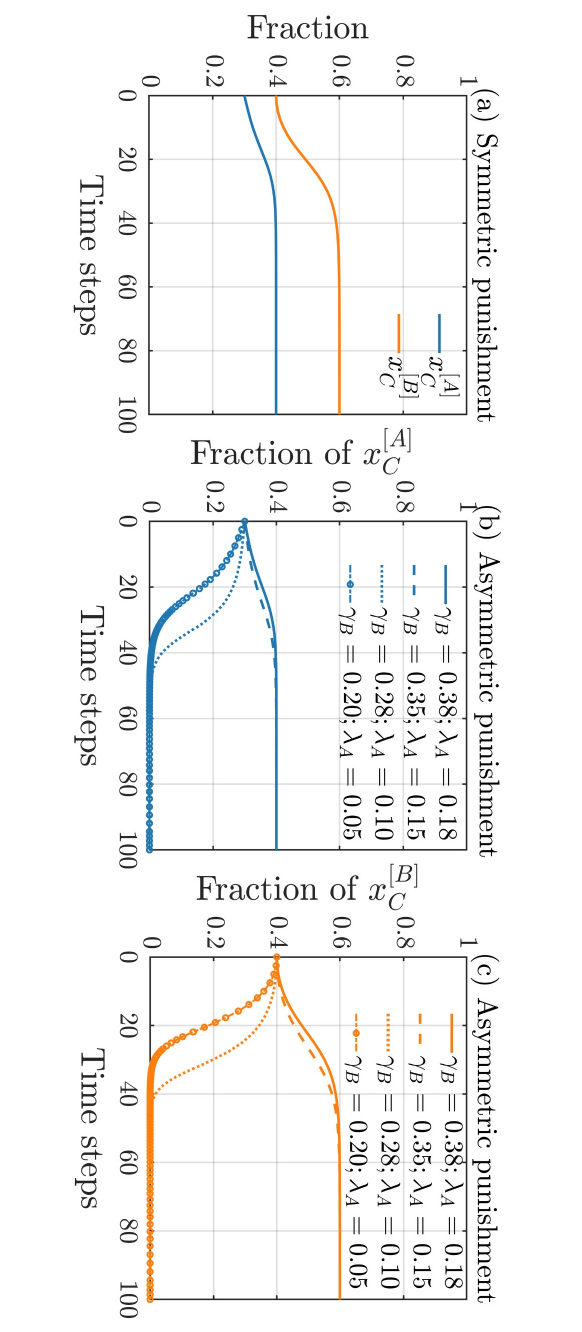}
		}
		\caption{\label{fig:7} Comparison of the time evolution of cooperator fractions for different subsets under symmetric and asymmetric punishment mechanisms. Panel~(a) corresponds to the symmetric punishment case with $\lambda_{A,B}=0.2$, $\gamma_{A,B}=0.4$. Panels~(b) and (c) show the asymmetric punishment case for four parameter pairs $(\gamma_B,\lambda_A)\in{(0.38,0.18),(0.35,0.15),(0.28,0.1),(0.2,0.05)}$. Other parameters are $N=5, r=3, \alpha=0.4, x_0=(0.3,0.4)$.}
	\end{figure*}
	
	When $\alpha$ increases ($\alpha = 0.6$), the threshold is primarily determined by the punishment intensity $\gamma_B$ exerted by the disadvantaged subset-$B$ individuals. As a result, a larger value of $r$ is required for the system to reach full cooperation, as shown in Fig.~\ref{fig:6}(c) and \ref{fig:6}(d). As the dominance of subset-$A$ cooperators diminishes, their enforcement influence weakens, while subset-$B$ cooperators gain relative leverage. The system requires stronger incentives (higher $r$) for cooperation to propagate effectively. Fig.~\ref{fig:6}(c) indicates that the absolute advantage of subset-$A$ individuals gradually diminishes as $r$ increases, and subset-$B$ individuals eventually surpass them at $r=4.2$. Asymmetric punishment between distinct subgroups creates a dynamic balance of influence: dominant subgroups can spearhead cooperation when they are prevalent, but the disadvantaged subset can assume responsibility only if sufficient incentives exist, making the propagation of cooperation contingent on both $r$ and the initial subset distribution. These results demonstrate that increasing $r$ universally promotes cooperation, but the relative influence of each subset on achieving stable cooperation depends critically on the initial equilibrium distribution $\alpha$.
	
	After presenting the effects of asymmetric punishment between distinct subsets on cooperation, it is necessary to compare them with the symmetric-punishment setting to obtain a complete picture of the punishment structure between distinct subsets. Fig.~\ref{fig:7} compares the evolution of cooperation rates over time for two subsets under symmetric vs asymmetric punishment mechanisms. Under symmetric punishment, cooperators in both subsets converge to an internal equilibrium point, as shown in Fig.~\ref{fig:7}(a), indicating that fair punishment sustains stable cooperation. This occurs because symmetric punishment ensures that cooperators in both subsets bear same costs and exert balanced enforcement, maintaining the relative payoff advantage of cooperative behavior and preventing either subset from being disproportionately disadvantaged.
	
	However, when the punishment mechanism becomes asymmetric, the attraction domain of the internal equilibrium point is significantly reduced. Asymmetric punishment creates imbalances in costs and enforcement power between the subsets, so that one subset may be overburdened, undermining the relative payoff of cooperators and destabilizing the cooperative equilibrium. As shown in Fig.~\ref{fig:7}(b), with an increasing degree of punishment asymmetry, the cooperation level of subset-$A$ can no longer remain near $\alpha$ and is eventually driven to the full defection state $x_C^{[A]}=0$. Fig.~\ref{fig:7}(c) indicates that subset-$B$ exhibits a similar trend. In conclusion, asymmetric punishment undermines the conditions necessary to sustain cooperation, eliminating the fully cooperative state. Under punishment between distinct subsets, cooperation can be maintained only when punishment is sufficiently fair; once the punishment becomes unfair, the system inevitably collapses into universal defection.
	
	\begin{figure*}[htbp]
		\hspace*{-0.2cm}
		\adjustbox{rotate=90, max width=0.6\textheight, center}{
			\includegraphics[trim=0.5cm 0cm 0.5cm 0cm, clip]{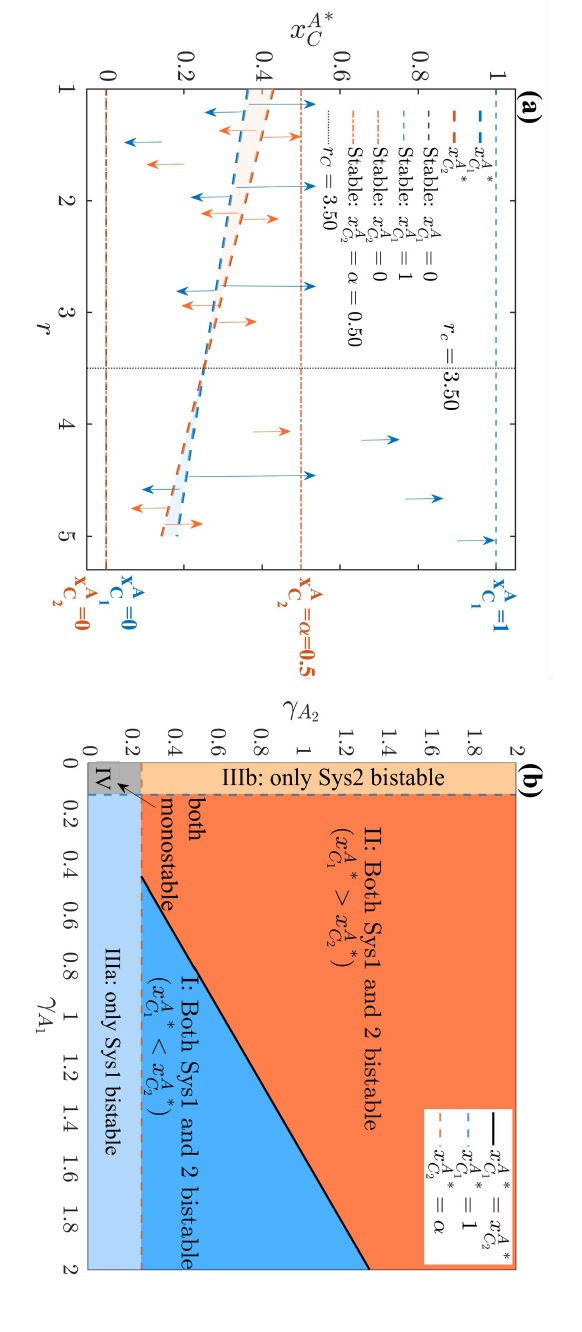}
		}
		\caption{\label{fig:8} The equilibria of the uniformly applied punishment (system~1, blue) and the selected punishment in the presence of distinct subsets (system~2, orange) as functions of synergy factor $r$ (panel~(a)) and punishment fines $\gamma_{A_1}$ and $\gamma_{A_2}$ (panel~(b)). (a) Arrows indicate the direction of evolutionary dynamics, with solid and dashed lines representing stable and unstable equilibria, respectively. As $r$ increases, the basin of attraction for full cooperation expands in both systems. (b) The orange and blue regions indicate bistability in both systems, while light orange and light blue represent bistability in system~2 or system~1 only. The gray area corresponds to full defection in both systems. The black solid line denotes $x_{C1}^{A*}=x_{C2}^{A*}$, with the region above (below) favoring full cooperation in system~2 (system~1). Dashed lines mark the boundary between monostable and bistable regimes. Overall, increasing the punishment intensity facilitates the emergence of cooperation. Other parameters are $N=5$, $\lambda_{A_1}=0.2$, $\lambda_{A_2}=0.2$, and $\alpha=0.5$. (a) $\gamma_{A_1}=0.9, \gamma_{A_2}=0.5$ and (b) $r=2.5$.}
	\end{figure*}
	
	Serving as the study’s final investigation, Fig.~\ref{fig:8} elucidates how within-subgroup and punishment between distinct subgroups mechanisms distinctly shape the evolution of cooperation, highlighting the nuanced understanding afforded by their comparative analysis. Here, the comparison between the two systems is made under the assumption of symmetric punishment between distinct subgroups. The results show that increasing $r$, $\gamma_{A_1}$, and $\gamma_{A_2}$ facilitates the evolution of cooperation in both systems. 
	
	When the enhancement factor $r$ is low ($r < r_C = 3.5$), the uniformly applied punishment (system~1) can reach the cooperative basin of attraction more easily, whereas the dedicated punishment in a bipartite population (system~2) encounters greater difficulty (Fig.~\ref{fig:8}(a)). The uniformly applied punishment protocol allows cooperators to suppress defectors at a lower threshold, thereby promoting cooperation. This occurs because 
	the collective efforts of cooperators allow
	defectors to be suppressed effectively even at low $r$, whereas selected punishment between distinct subsets initially disperses enforcement, requiring higher collective incentives to achieve similar suppression. As $r$ increases, the advantage of system~2 becomes apparent. The broader scope of punishment between distinct subsets enables cooperators to more effectively restrain defectors in mixed groups, yielding a lower threshold for entering the cooperative basin compared with system~1. This indicates that at high enhancement factors, punishment between distinct subsets is more effective in stabilizing cooperation. Overall, system~1 is more advantageous at low enhancement factors, whereas system~2 becomes more effective at high enhancement factors.
	
	As shown in Fig.~\ref{fig:8}(b), when the enhancement factor is relatively low ($r = 2.5$), system~2 is generally more effective than system~1 at promoting cooperation across a wide range of punishment intensities. Yet, with increasing $\gamma_{A_1}$ and $\gamma_{A_2}$, the advantage of system~2 gradually diminishes, while that of system~1 strengthens. At moderate punishment levels, punishment between distinct subsets exerts a widespread influence across the sets, with cooperators in one subset reaching into the other subset to restrain defectors, amplifying the overall cooperative influence. As punishment intensity increases, each subset forms a “fortified bastion” of local enforcement, where defectors are tightly contained within their own subset, making the added benefit of cross-subset interactions less significant. Notably, the gray region in Fig.~\ref{fig:8}(b) represents full defection in both systems, highlighting that insufficient punishment intensity cannot sustain cooperation, regardless of the mechanism. 
	
	\begin{figure*}
		\centering
		\includegraphics[width=0.8\linewidth,
		trim=0cm 5.0cm 0cm 5.0cm, clip]{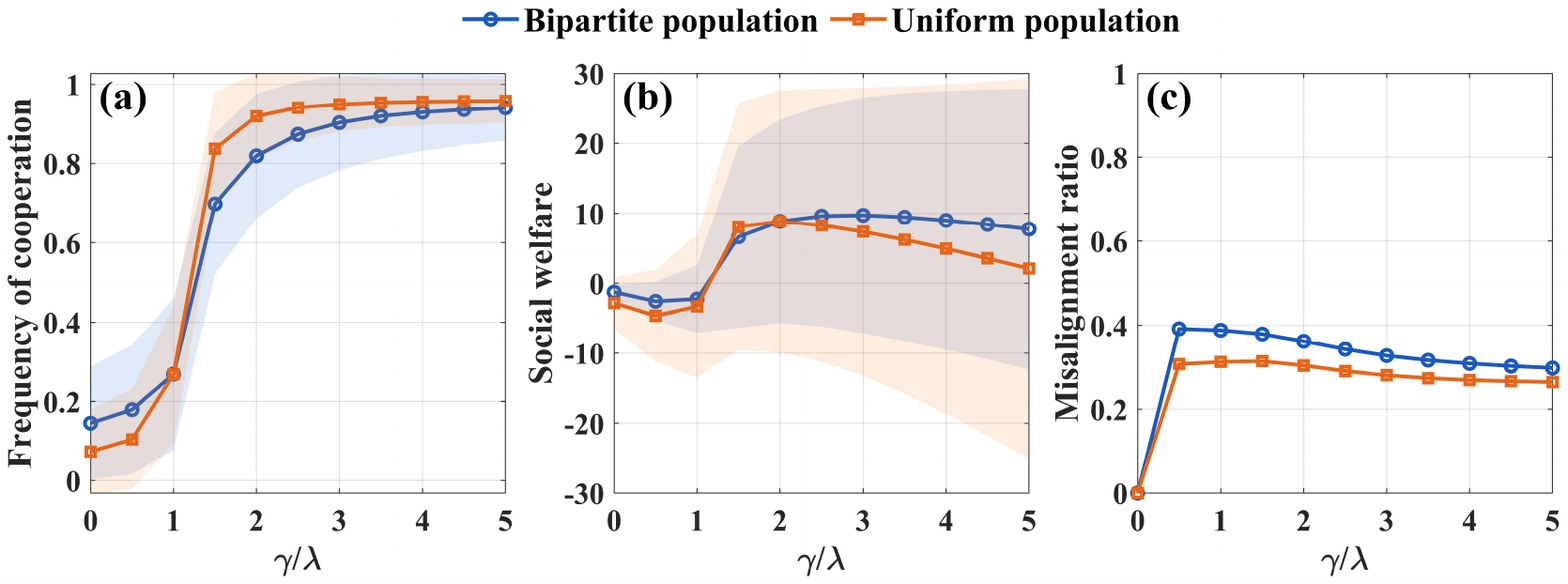}
		\caption{The impact of punishment on cooperation and social welfare, for varying enforcement-to-cost ratio of incentives $\gamma/\lambda$. (a,b) Impact of punishment for cooperation and social welfare. Shown is the frequency of cooperation and social welfare between bipartite and uniform populations. (c) The impact of punishment for misalignment. The fraction of misalignment as a function of the cost ratio of punishment $\gamma/\lambda$ among different population is shown. The parameter space is $\mu\in\{0.001,0.01,0.1\}$, $\beta\in\{0.1,1\}$, $\gamma\in\{0,5\}$, $\lambda=1$. The orange and blue lines represent the average results for the uniform population and the bipartite population, respectively. The orange and blue shading indicates standard deviation.}
		\label{fig:9}
	\end{figure*}
	
	Taken across the entire range of punishment intensities, punishment between distinct subsets remains superior overall because of its ability to coordinate cooperative enforcement across subsets ensures that even at moderate or low intensities, defectors are constrained more effectively than with local enforcement alone, providing a more robust pathway to maintain cooperation under diverse conditions. Overall, these results suggest that cooperators in system~2 are particularly effective at fostering cooperation when applying relatively weak punishment to defectors in distinct subsets, while system~1 dominates when punishment is stronger or local enforcement suffices.
	
	Following the comparison of cooperation levels, we further investigate how punishment intensity and cost ratios influence social welfare and the alignment between cooperation and collective outcomes in finite populations, using the analytical framework proposed by Han~\cite{han26} et al. We consider two population structures, namely, a bipartite population and a uniform population, and systematically examine the evolutionary outcomes as the punishment impact-to-cost ratio varies.
	
	Consistent with the work of Han et al.,~\cite{han26} the level of cooperation is quantified by the stationary frequency of cooperators. For a given parameter configuration, the frequency of strategy $i$, in particular cooperation, is obtained by averaging over all possible population states 
	$\mathbf{n}$, weighted by the corresponding stationary distribution $\bar{p}_{\mathbf{n}}$~\cite{lorenz05},
	\begin{equation}
		f_i = \sum_{\mathbf{n}} \frac{n_i}{N} \bar{p}_{\mathbf{n},}
	\end{equation}
	where $n_i$ denotes the number of individuals adopting strategy $i$ in state $\mathbf{n}$ and $N$ is the population size.
	
	Similarly, social welfare is defined as the expected total population payoff under the stationary distribution,
	\begin{equation}
		SW = \sum_{\mathbf{n}} SW(\mathbf{n}) \bar{p}_{\mathbf{n},}
	\end{equation}
	where $SW(\mathbf{n})$ represents the total payoff of the population when it is in state $\mathbf{n}$. These definitions ensure that both cooperation and social welfare capture the long run behavior of the stochastic evolutionary process.
	
	However, higher levels of cooperation do not necessarily imply higher social welfare. To capture this potential tension, we further examine the alignment between cooperation and social welfare as the punishment impact-to-cost ratio varies. Therefore, alignment is assessed by comparing the directional changes of the cooperation frequency $f_C(u)$ and social welfare 
	$SW(u)$ with respect to a control parameter $u$, which in this study corresponds to the punishment impact-to-cost ratio. For two successive parameter values $u_k$ and $u_{k+1}$, the forward differences are computed as $\Delta f_C(u_k) = f_C(u_{k+1}) - f_C(u_k)$ and
	$\Delta SW(u_k) = SW(u_{k+1}) - SW(u_k)$. The two objectives are classified as aligned if these differences share the same sign, and as misaligned otherwise,
	\begin{equation}
		\text{Misalignment} \iff \operatorname{sgn}\!\left(\Delta f_C(u_k)\right) \neq \operatorname{sgn}\!\left(\Delta SW(u_k)\right),
	\end{equation}
	where $\operatorname{sgn}(x)$ denotes the sign function, which takes the value 1 when $x>0$, 0 when $x=0$, and $-1$ when $x<0$.
	
	Using the above measures, we present in Fig.~9 the effects of the punishment impact-to-cost ratio on cooperation, social welfare, and their alignment for bipartite and uniform populations. Fig.~9(a) shows the effect of punishment on cooperation with $\gamma/\lambda$. For small values of this ratio, the bipartite population exhibits a higher level of cooperation than the uniform population, indicating that population heterogeneity enhances the effectiveness of weak punishment. However, as the punishment impact-to-cost ratio increases, this advantage gradually disappears and eventually reverses. For sufficiently large ratios, the uniform population attains higher cooperation levels, suggesting that strong punishment is more efficiently exploited in homogeneous populations.
	
	Fig.~9(b) shows that social welfare depends nonmonotonically on the punishment impact-to-cost ratio: it first decreases, then reaches an intermediate maximum, and finally declines at high ratios. The bipartite population consistently achieves higher social welfare than the uniform population, indicating that heterogeneity can buffer welfare losses. Fig.~9(c) shows a similar nonmonotonic pattern for the misalignment between cooperation and social welfare. Misalignment is highest at intermediate punishment levels and decreases when punishment is very weak or very strong. The bipartite population consistently exhibits a higher misalignment than the uniform population. In summary, Fig.~9 shows that population heterogeneity increases social welfare while also amplifying the gap between cooperation and collective outcomes.

	\section{Discussion}
	In the real world, punishment between distinct groups is ubiquitous~\cite{frederickson13}. In nature, diverse forms of reciprocal punishment are observed~\cite{gauduchon2002cheating}. In human societies, similar regulatory behaviors occur between different organizations or groups~\cite{gelfand24}. Within teams, supervision and punishment take place between superiors and subordinates~\cite{tepper17}. However, most studies on punishment have been limited to homogeneous systems. 
	
	Building on the public goods game framework, this study examines how selected punishment between distinct subsets shapes cooperative dynamics in contrast to a uniformly applied punishment. We find that cooperation can emerge and remain stable when punishment between distinct subsets is symmetric. Such symmetry balances enforcement across subsets, leaving no position exposed to exploitation and allowing cooperative influence to spread through the population. In particular, at relatively low punishment intensity and enhancement levels, symmetric punishment between distinct subsets outperforms uniformly applied punishment. Under these conditions, cooperators can project enforcement across subsets, letting cooperative pressure exert a widespread influence across the subsets, whereas uniform punishment keeps enforcement confined and limits its reach. A further interesting observation is that, in bipartite populations, cross-subset punishment increases social welfare and amplifies the gap between cooperation and collective outcomes. 
	
	Notably, the results are not entirely intuitive. While cooperators in the bipartite population pay lower punishment costs, and defectors receive correspondingly less punishment, the observed enhancement of cooperation arises from the structure of cross-subset enforcement rather than from reduced costs alone. Once this symmetry is disrupted, however, cooperation declines. As enforcement becomes uneven, some subsets become weak points--either overloaded or insufficiently protected--reducing the advantage of cooperating and giving defectors opportunities to invade. This imbalance ultimately breaks the coherence of the enforcement structure and undermines stable cooperation.
	
	These patterns are consistent with evidence from behavioral economics, where symmetric and reciprocal sanctioning opportunities lead to higher cooperation than unilateral or group-internal punishment~\cite{fehr02,nikiforakis08}. As well as with real-world observations from hierarchical organizations, where asymmetric punishment structures make subordinates more likely to withdraw cooperation first, thereby undermining collective performance ~\cite{near95,colnaghi25}. Similar dynamics appear in biological mutualisms: in ant–acacia systems, cooperation is stabilized not by within-species sanctions but by cross-species reciprocal enforcement, where plants reduce rewards to poorly defending ants and ants, in turn, impose costs on low-reward plants~\cite{palmer07,palmer13}. In both human and biological systems, symmetry and coordinated enforcement act like an invisible scaffold, channeling incentives and preventing exploitation, which explains why cooperative behavior persists under balanced punishment between distinct subsets. 
	
	These interdisciplinary findings highlight a common principle: symmetry and coordination in punitive relationships, rather than sanction magnitude alone, determine whether punishment enhances or undermines cooperation. The results here extend this principle to bipartite social populations, showing that symmetric punishment between distinct subsets can generate a more integrated enforcement structure than a uniform punishment. However, once this symmetry breaks, punishment between distinct subsets may instead destabilize cooperation that would otherwise persist.
	
	In light of the analysis outlined above, the analysis suggests that the effectiveness of punishment-based interventions depends not only on their presence, but on how key parameters governing enforcement and cooperative benefits are jointly configured.
	\begin{itemize}[left=0pt, topsep=0pt, itemsep=0pt, parsep=0pt]
		\item In multi-group or hierarchical settings, managers should calibrate the strength of sanctions ($\gamma$) so that no group is disproportionately burdened. This can be achieved by monitoring enforcement loads and scaling punishment for weaker groups, or by implementing coordinated cross-group enforcement where cooperative individuals in one group can influence the behavior of another. Such symmetry prevents defection cascades caused by overburdened groups.
		\item High sanctioning costs ($\lambda$) in a bipartite population can destabilize cooperation, as the subset bearing a disproportionate share of enforcement may collapse under asymmetric punishment. To prevent this, managers can balance the enforcement burden by allowing cooperative individuals in the less burdened subset to assist in punishing defectors, sharing monitoring responsibilities, or providing targeted incentives to reduce punishment costs. Coordinating costs in this way helps maintain cooperation across both subsets and prevents defection cascades.
		\item When the collective enhancement $r$ of cooperation is low, uniform, localized punishment may suffice, allowing cooperators to contain defectors efficiently within each subgroup. As the potential cooperative benefits increase, targeted cross-group punishment becomes more effective, enabling cooperators to suppress defectors across subsets. Policy should, therefore, adapt the enforcement scheme to the expected payoff structure, using uniform punishment for low-return scenarios and cross-group punishment for high-return cooperation contexts.
		\item In multi-group or hierarchical settings, coordinating punishment across subgroups can more effectively enhance overall social welfare than uniform or localized enforcement. Cross-group accountability distributes enforcement more evenly, reduces weak points, and strengthens collective outcomes.
	\end{itemize}

	Future work could extend our framework by considering spatially structured populations or genuinely bipartite network topologies, where interactions are constrained by network links rather than purely by labels. Such extensions would allow exploration of how local interaction patterns and network architecture influence the effectiveness of intergroup punishment and the emergence of cooperation.
	
	\section*{Acknowledgments}
	This work is supported by the National Key R\&D Program of China (Grant No. 2021YFA1000402), National Natural Science Foundation of China (Grant No. 72571215), Guangdong Basic and Applied Basic Research Foundation (Grant No. 2024A1515011241), Key Scientific Research Program Project of Shaanxi Provincial Department of Education (Grant No. 24JS055), the Natural Science Basic Research Plan in Shaanxi Province of China (Grant No. 2024JC-YBMS-588), and by the National Research, Development and Innovation Office (NKFIH), Hungary under Grant No. K142948.
	
	\section*{AUTHOR DECLARATIONS}
	\subsection*{Conflict of Interest}
	The authors have no conflicts of interest to disclose.
	\subsection*{Author Contributions}
	\textbf{Sinan Feng}: Conceptualization (equal); Formal analysis (equal); Methodology (equal); Project administration (equal); Visualization (lead); Writing - original draft (lead); Writing - review \& editing (equal). \textbf{Genjiu Xu}: Project administration (equal); Supervision (equal); Writing - review \& editing (equal). \textbf{Yu Chen}: Writing - review \& editing (equal). \textbf{Chaoqian Wang}: Conceptualization (lead); Formal analysis (equal); Methodology (equal); Project administration (equal); Validation (equal); Visualization (equal); Writing - review \& editing (equal). \textbf{Attila Szolnoki}: Project administration (equal); Validation (equal); Visualization (equal); Writing - review \& editing (equal). 
	
	\section*{DATA AVAILABILITY}
	The data that support the findings of this study are available from the corresponding author upon reasonable request.
	
	\appendix
	
	\section{Replicator dynamics in uniform population}
	
	When punishment occurs only within the same group, we analyze separately the effect of punishment on cooperation. First, the replicator equation for the group is:
	\begin{equation}
		\begin{cases}
			\dot{x}_C =  \displaystyle{x_C \bigl(1 - x_C\bigr)\left[\frac{r}{N} - 1 - \lambda_A (N - 1)x_D + \gamma (N - 1)x_C \right]},
			\\[1em]
			\dot{x}_D =  \displaystyle{x_D \bigl(1 - x_D\bigr)\left[-\frac{r}{N} + 1 - \gamma (N - 1)x_C + \lambda (N - 1)x_D \right]}.
		\end{cases}
	\end{equation}
	
	By solving the above differential equation, three equilibrium points can be obtained: $(1,0), (0,1), (\displaystyle{\frac{(N - 1)\lambda + 1 - r/N}{(N - 1)(\gamma + \lambda)}}
	,\displaystyle{\frac{(N - 1)\gamma - 1 + r/N}{(N - 1)(\gamma + \lambda)})}$. The Jacobian matrices for the three equilibrium points are:
	
	\[
	J{(1,0)} =
	\begin{pmatrix}
		- \left(\dfrac{r}{N} - 1 + \gamma (N - 1)\right) & 0 \\[6pt]
		0 & -\gamma (N - 1) - \dfrac{r}{N} + 1
	\end{pmatrix},
	\]
	\[
	J{(0,1)} =
	\begin{pmatrix}
		\dfrac{r}{N} - 1 - \lambda (N - 1) & 0 \\[6pt]
		0 & -\left( 1 - \dfrac{r}{N} + \lambda (N - 1) \right)
	\end{pmatrix},
	\]
	and
	\[
	J{\left((x_C)^*=
		\displaystyle{\frac{(N - 1)\lambda + 1 - \displaystyle{\frac{r}{N}}}{(N - 1)(\gamma + \lambda)}} \; , \;
		(x_D)^*=\displaystyle{\frac{(N - 1)\gamma - 1 + \displaystyle{\frac{r}{N}}}{(N - 1)(\gamma + \lambda)}}
		\right)}=\]
	\[
	\begin{pmatrix}
		(x_C)^{*}\bigl(1 - (x_C)^{*}\bigr)\,\gamma (N - 1) 
		& -(x_C)^{*}\bigl(1 - (x_C)^{*}\bigr)\,\lambda (N - 1) \\[6pt]
		-(x_D)^{*}\bigl(1 - (x_D)^{*}\bigr)\,\gamma (N - 1) 
		& (x_D)^{*}\bigl(1 - (x_D)^{*}\bigr)\,\lambda (N - 1)
	\end{pmatrix}.
	\]
	
	Next, we analyze the stability of each equilibrium point.\\
	1. For $(1,0)$, the eigenvalues of the Jacobian matrix are $\displaystyle{\lambda_1=-\left( r/N - 1 + \gamma (N - 1) \right)}$,\\ $\displaystyle{\lambda_2=-\gamma (N - 1) - r/N + 1}$. When $r \geq N(1-\gamma (N-1))$, both of which are negative, indicating that $(1,0)$ is a stable node.\\
	2. For $(0,1)$, the eigenvalues of the Jacobian matrix are $\displaystyle{\lambda_1=r/N - 1 - \lambda (N - 1)}$, $\displaystyle{\lambda_2=-\left( 1 - r/N + \lambda (N - 1) \right)}$. Both of which are negative, indicating that $(0,1)$ is a stable node.\\
	3. For $\displaystyle{(\frac{(N - 1)\lambda + 1 - r/N}{(N - 1)(\gamma + \lambda)}}
	,\frac{(N - 1)\gamma - 1 + r/N}{(N - 1)(\gamma + \lambda)})$, the condition for the existence of this equilibrium point is $r < N \left[ (N - 1)\lambda + 1 \right]$ and $r > N \left[ 1 - (N - 1)\gamma \right]$. The eigenvalues of the Jacobian matrix at this point are $\lambda_1=0$, $\lambda_2=\frac{\bigl[N \gamma (N - 1) + (r - N)\bigr] \, \bigl[N \lambda (N - 1) - (r - N)\bigr]}{N^2 (N - 1)(\gamma + \lambda)}$. However, when $r > N \left[ 1 - (N - 1)\gamma \right]$, $\lambda_2>0$. So, this point is unstable.
	
	\section{Replicator Dynamics in bipartite population}
	When punishment can be enforced across subgroups, let the proportion of individuals with subgroup-$A$ be denoted by $\alpha$, and the proportion of individuals with subgroup-$B$ by $1-\alpha$. The replicator dynamics of the system is given by
	\begin{equation}
		\begin{cases}
			\begin{aligned}
				\dot{x}_{C}^{[A]} 
				&= \displaystyle{x_{C}^{[A]}\bigl(\alpha - x_{C}^{[A]}\bigr)\bigl(f_{C}^{[A]} - f_{D}^{[A]}\bigr)} \\[4pt]
				&= \displaystyle{x_{C}^{[A]}\bigl(\alpha - x_{C}^{[A]}\bigr)
					\Bigl[
					\frac{r}{N} - 1 
					- (N - 1)\lambda_{A}\bigl(1 - \alpha - x_{C}^{[B]}\bigr)} \\[-2pt]
				&
				\displaystyle{+ (N - 1)\gamma_{B}x_{C}^{[B]}
					\Bigr]}, \\
				\dot{x}_{C}^{[B]} 
				&= \displaystyle{x_{C}^{[B]}\bigl(1 - \alpha - x_{C}^{[B]}\bigr)\bigl(f_{C}^{[B]} - f_{D}^{[B]}\bigr)}\\[4pt]
				&= \displaystyle{x_{C}^{[B]}\bigl(1 - \alpha - x_{C}^{[B]}\bigr)
					\Bigl[
					\frac{r}{N} - 1 
					- (N - 1)\lambda_{B}\bigl(\alpha - x_{C}^{[A]}\bigr)} \\[-2pt]
				&
				+ (N - 1)\gamma_{A}x_{C}^{[A]}
				\Bigr].
			\end{aligned}
		\end{cases}
	\end{equation}
	
	There are five equilibrium points: 
	$(0,0)$, $(0,1-\alpha)$, $(\alpha,0)$, $(\alpha,1-\alpha)$, \\
	$\left(\displaystyle{\frac{1-\displaystyle{\frac{r}{N}}+(N-1)\lambda_B\alpha}{(N-1)(\gamma_A+\lambda_B)}},\displaystyle{\frac{1-\displaystyle{\frac{r}{N}}+(N-1)\lambda_A(1-\alpha)}{(N-1)(\lambda_A+\gamma_B)}}\right)$.
	
	The Jacobian matrices for these equilibrium points are:
	\begin{widetext}
		\begin{equation*}
			\begin{aligned}
				J(0,0)
				&=
				\begin{pmatrix}
					\displaystyle
					\alpha \Bigl(\frac{r}{N} - 1 - (N-1)\lambda_A (1-\alpha)\Bigr) & 0 \\
					0 & \displaystyle (1-\alpha) \Bigl(\frac{r}{N} - 1 - (N-1)\lambda_B \alpha\Bigr)
				\end{pmatrix},\\[5pt]
				J\bigl(0,\,1-\alpha\bigr)
				&=
				\begin{pmatrix}
					\displaystyle
					\alpha \Bigl[\frac{r}{N} - 1 + (N-1)\gamma_B (1-\alpha)\Bigr] & 0 \\
					0 & \displaystyle -(1-\alpha) \Bigl[\frac{r}{N} - 1 - (N-1)\lambda_B \alpha\Bigr]
				\end{pmatrix},\\[5pt]
				J(\alpha,0)
				&=
				\begin{pmatrix}
					-\alpha \Bigl[\dfrac{r}{N} - 1 - (N-1)\lambda_A (1-\alpha)\Bigr] & 0 \\
					0 & (1-\alpha) \Bigl[\dfrac{r}{N} - 1 + (N-1)\gamma_A \alpha\Bigr]
				\end{pmatrix},\\[5pt]
				J(\alpha,1-\alpha)
				&=
				\begin{pmatrix}
					-\alpha \Bigl[\dfrac{r}{N} - 1 + (N-1)\gamma_B (1-\alpha)\Bigr] & 0 \\
					0 & -(1-\alpha) \Bigl[\dfrac{r}{N} - 1 + (N-1)\gamma_A \alpha\Bigr]
				\end{pmatrix},\\[5pt]
				J\Bigl((x_C^{[A]})^*, (x_C^{[B]})^*\Bigr)
				&=
				J\Biggl(
				\frac{1-\frac{r}{N}+(N-1)\lambda_B\alpha}{(N-1)(\gamma_A+\lambda_B)},\,
				\frac{1-\frac{r}{N}+(N-1)\lambda_A(1-\alpha)}{(N-1)(\lambda_A+\gamma_B)}
				\Biggr)\\[5pt]
				&=
				\begin{pmatrix}
					0 & (x_C^{[A]})^* (\alpha - (x_C^{[A]})^*) (N-1)(\lambda_A + \gamma_B) \\
					(x_C^{[B]})^* (1-\alpha - (x_C^{[B]})^*) (N-1)(\lambda_B + \gamma_A) & 0
				\end{pmatrix}.
			\end{aligned}
		\end{equation*}
	\end{widetext}

	Next, we analyze the stability of each equilibrium point.\\
	1. For $(0,0)$, the eigenvalues of the Jacobian at this point are both negative, i.e., $\alpha \Bigl[ r/N - 1 - (N-1)\,\lambda_A (1-\alpha) \Bigr]<0$ and $(1-\alpha) \Bigl[ r/N - 1 - (N-1)\, \lambda_B \alpha \Bigr]<0$. Therefore, $(0,0)$ is a stable node.\\
	2. For $(0,1-\alpha)$, the eigenvalues of the Jacobian at this point are $\alpha \Bigl[ r/N - 1 + (N-1)\,\gamma_B (1-\alpha) \Bigr]$ and $-(1-\alpha)\Bigl[r/N - 1 - (N-1)\,\lambda_B \alpha\Bigr]$. When $r<N\Bigl[1-(N-1)\gamma_B(1-\alpha)\Bigr]$, this point is a saddle point.\\
	3. For $(\alpha,0)$, given that $r<N\Bigl[1-(N-1)\gamma_A\alpha\Bigr]$, the eigenvalue of the Jacobian at this point is a positive and a negative, i.e.,\\
	\[\displaystyle{-\alpha \Bigl[\frac{r}{N} - 1 - (N-1)\,\lambda_A (1-\alpha)\Bigr]}>0\]\\
	and\\
	\[
	(1-\alpha)\Bigl[\frac{r}{N} - 1 + (N-1)\,\gamma_A \alpha\Bigr]<0.
	\]\\
	
	Therefore, $(\alpha,0)$ is a saddle point.\\
	4. Consider the equilibrium $(\alpha,1-\alpha)$, given $r>max\Biggl\{N\Big[1-(N-1)\gamma_B(1-\alpha)\Big],N\Big[1-(N-1)\gamma_A\alpha\Big]\Biggr\}$, i.e., $
	-\alpha \Bigl\{ r/N - 1 + (N-1)\,\gamma_B (1-\alpha) \Bigr\}$and $-(1-\alpha)\Bigl[r/N - 1 + (N-1)\,\gamma_A \alpha\Bigr]$, the eigenvalues of the Jacobian at this point are both negative. So, this point is a stable node. \\
	5. For $\left(
	\displaystyle{\frac{\,1 - \displaystyle{\frac{r}{N}} + (N-1)\lambda_B \alpha\,}{(N-1)(\gamma_A + \lambda_B)},\;
		\frac{\,1 - \displaystyle{\frac{r}{N}} + (N-1)\lambda_A (1-\alpha)\,}{(N-1)(\lambda_A + \gamma_B)}
	}\right)$, the corresponding two eigenvalues are \\
	\begin{align*}
		& (N-1)\sqrt{(\lambda_A+\gamma_B)(\lambda_B+\gamma_A)
			(x_C^{[A]})^*(\alpha-(x_C^{[A]})^*)} \\
		& \qquad \times \sqrt{(x_C^{[B]})^*(1-\alpha-(x_C^{[B]})^*)} > 0
	\end{align*}
	and
	\begin{align*}
		& -(N-1)\sqrt{(\lambda_A+\gamma_B)(\lambda_B+\gamma_A)
			(x_C^{[A]})^*(\alpha-(x_C^{[A]})^*)} \\
		& \qquad \times \sqrt{(x_C^{[B]})^*(1-\alpha-(x_C^{[B]})^*)} < 0.
	\end{align*}
	
	Therefore, this internal equilibrium point is unstable and is an internal saddle point.

	\nocite{*}
	
	%

\end{document}